\documentclass{elsart}
\usepackage{graphicx}
\newcommand{\Vec}[1]{\mbox{\boldmath{$#1$}}}
\newcommand{\Nab}{\Vec{\nabla}}
\newcommand{\Dif}[2]{\frac{\partial{}#1}{\partial{}#2}}
\newcommand{\Unit}[1]{\,\mbox{$\rm #1$}}
\newcommand{\eV}{\mbox{eV}}
\begin{document}
\runauthor{Schopper, Birk and Lesch}
\begin{frontmatter}
\title{High Energy Hadronic Acceleration in Extragalactic Radio Jets}
\author{R\"udiger Schopper\thanksref{label1}}
\author{, Guido Thorsten Birk}
\author{and Harald Lesch}
\address{Institut f{\"u}r Astronomie und Astrophysik, Universit{\"a}t
M{\"u}nchen, Germany\\
Centre for Interdisciplinary Plasma Science~(CIPS), Garching, Germany}
\thanks[label1]
{Institut f{\"u}r Astronomie und Astrophysik, Universit{\"a}t
M{\"u}nchen,
Scheinerstra{\ss}e 1, D-81679 M{\"u}nchen, Germany\\
e-mail: schopper@usm.uni-muenchen.de\\
phone: +49 89 2180 6005\\
fax: +49 89 2180 6003}
\begin{abstract}
Within an electric circuit description of extragalactic jets temporal variations
of the electric currents are associated with finite collisionless conductivities
and consequently magnetic-field aligned electric fields $E_\parallel$.  The
maximum field strengths depend on the efficiency of the jet MHD generator
$E_\perp$ and the local conversion to the $E_\parallel$ component.  The hadronic
jet constituents can efficiently be accelerated in such fields all along the
jets.  To estimate the maximum energy the accelerated jet hadrons can achieve we
consider energy loss processes as photon-pion and pair production as well as
synchrotron and inverse Compton radiation.  It turns out that for the strongest
$E_\parallel$ possible the Centaurus A jet is a most promising candidate for the
source of the highest energy component of cosmic rays.
\end{abstract}
\end{frontmatter}

\section{Introduction}
Extragalactic jets are nonthermal radiative and collimated magnetised plasma
flows which predominantly consist of relativistic particles and magnetic fields.
They originate from the very centres of active galaxies and dissipate most of
their energies in hotspots and lobes far away from the host galaxy.  Jet lengths
range from a few kpc up to 100 kpc.  The jet composition is not yet clear, since
their nonthermal radiation (mainly synchrotron emission) provides information
only about the leptonic constituents.  The nonthermal radiation output of the
jets extends from the radio even to the X-ray regime [1-3].  To emit X-rays via
the synchrotron process, the radiating leptons must have energies of several
\Unit{T\eV} or Lorentz factors up to $10^{7-8}$ [4].  These remarkably high
energy levels raise the problem of continuous re-acceleration simply because the
particle's synchrotron loss lengths are orders of magnitudes smaller than the
jet lengths along which the radiation is observed.  Especially the almost
constant radiation spectral index along the jets requires an almost constant
lepton energy that closely follows the behaviour of the local magnetic energy
density all along the jet [5].

The fact that nonthermal optical and X-ray emission is observed in jets means
that the charged particles must experience a continuous energy gain up to
\Unit{T\eV}-energies.  This is only possible by field aligned electric fields.
Especially the constancy of the spectral index, varying only together with the
local energy density of the magnetic field, give strong constraints for possible
acceleration models.  In particular, shock acceleration scenarios, in which also
electric fields are the ultimate cause for particle acceleration, have serious
problems explaining the astonishing continuous spectral index [5].  This constancy
shows, that there is a need for an interplay of a very efficient electromagnetic
energy gain and the radiative energy loss along the jets.  For leptons to fulfil
these requirements accelerating electric fields must be present all along the
path of the radiating jet.  This seems to be a trivial statement, however it has
profound implications for the hadrons in the jets, which are not detectable by
their radiational output.

This leads to the subject of the jet material: What kind of plasma is inside the
extragalactic jets? Are they made up of protons and electrons, or
electron-positron pairs, or a mixture of both? Discussing this important topic,
Celotti \& Fabian [6] argued in favour of proton-electron jets, whereas recently
on the grounds of circular polarisation measurements of some bright quasars pure
electron-positron jets [7, 8] have been suggested.  Sikora and Madejski [9]
derived strong constraints on the content of jets in quasars.  Their conclusion
was that the pair content of jets is high, but that dynamically the jets are
dominated by protons.  Especially they show that pure electron-positron jet
models overpredict soft X-ray emission, whereas pure proton-electron jet models
can be excluded since they predict too weak nonthermal X-radiation.  Jets may
dominantly consist leptons, but since the jets are supposed to be fed by
accretion disks consisting of normal electron-proton-plasmas, there is always a
hadronic component in extragalactic jets.  However, in our context there is no
need to precisely specify the relative abundance of the hadronic and leptonic
jet components, because here we want to analyse the energisation of individual
hadrons, only.

Before we dwell on the acceleration of individual particles within the jets, we
first describe the possible origin of the jets.  It is now generally agreed that
there is a link between the central activity of an active galactic nucleus, the
jets and the hotspots.  The basic idea that underlies the jet models is that a
jet is continuous collimated outflow which transports energy from the nucleus to
the hotspot [e.g.~10].  Many jets exhibit superluminal motions, i.e.~the plasma
bulk speed in the jet has to be close to the velocity of light with minimal
Lorentz factors between 2 and 10 [e.g.~11].  The only mechanism able to produce
these relativistic jets is the magnetic sling effect in a rapidly rotating
magnetosphere.  The collimation of the outflow is provided by the pinching effect
of the toroidal magnetic field.  This would resolve the puzzle that some jets are
not freely expanding and that an agent is necessary to confine them.
Additionally, the dynamical effects of the magnetic field ranges from refocusing
and knot formation to the possibility of electromagnetic interaction of the jets
with their surroundings.  One of the most important properties of jet production
in rotating magnetospheres is that jets carry an electrical current which flows
along the jet axis, a property which enhances their stability [12, 13].  The
source of the jet current is the differential rotation of the magnetosphere in
the innermost region of the galactic nucleus.  In order to avoid charging up the
nucleus a return current flowing in the surrounding plasma, must close the
circuit.  A current flow in and outside the jet requires current closure at the
tip of the jet where Ohmic dissipation by nonlinear collisionless plasma
processes may lead to particle acceleration and plasma heating in the hot spots
[14].

Now we have all the necessary ingredients for an electrodynamic model of
extragalactic jets: There is a shear flow power supply which provides a current
source associated with sheared magnetic fields in the jet.  The jet itself
represents the conductor along which the electric current flows.  Finally, at
least one well defined dissipation region can be identified- the hot spot and
lobe region that corresponds to a resistance.  However, the hotspot is not the
only dissipative region.  Since extragalactic jets show continously distributed
synchrotron emission continuous dissipation of particle kinetic energy into
electromagnetic waves occurs in the completely collisionless jet plasma.  The
non-ideal current flow is associated (via Ohm's law) with electric fields
oriented along the current direction and in the context of extragalactic jets
along the dominant poloidal magnetic field component.  The behaviour of this
electric field is a direct consequence of the self consistency of Amp\`ere's
Law.  Any variation in the current density must correspond to a change in the
magnetic shear (the energy source) and the aligned electric field.  It is this
property of the magnetic field--current system to resist any disturbance given
by radiative losses that provides the accelerating electric field.

Here we would like to address the question what it means for the hadronic jet
component, if such magnetic field-aligned electric fields are present in the jet
flows in the presence of the synchrotron radiation emitted by relativistic
leptons.  Whereas leptons suffer synchrotron losses that limit their energy gain
to a few \Unit{T\eV}, protons undergo additionally energy losses by pair and
pion production.  Nevertheless, their equilibrium energy (where energy gain and
energy loss processes are balanced) is significantly higher than the maximum
energy of the leptons.  We will show in this contribution that, in principle,
field-aligned electric fields in extragalactic jets can accelerate protons even
up to the highest energies measured for ultra-high-energy cosmic rays (UHECR)
$10^{21}$ \eV [15].
\section{Acceleration of Cosmic Ray Protons in the Electric Fields of
         Extragalactic Jets}
\begin{figure}
  \includegraphics[width=0.99\textwidth]{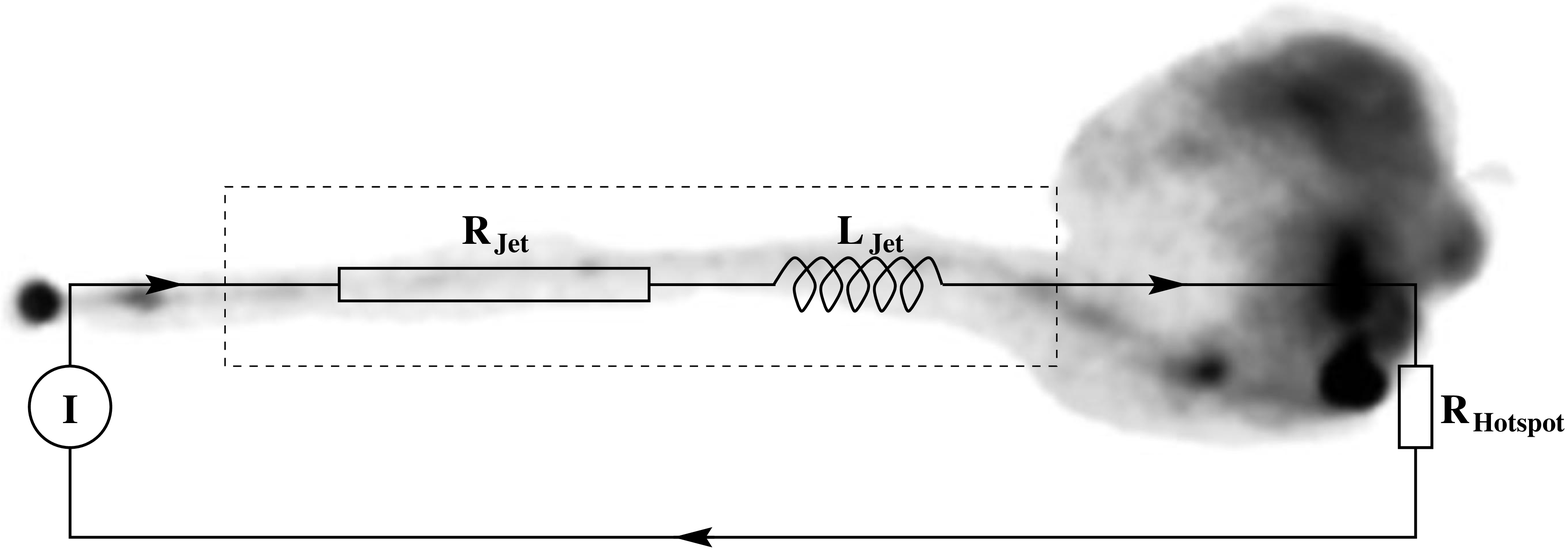}
  \caption{Illustration of the electric circuit that is equivalent to an
  extragalactic jet.  The current source~$I$ stands for the permanent magnetic
  shear injection of the Keplerian accretion disk surrounding the central
  supermassive black hole.  The underlying $6 \Unit{cm}$ radio image shows, for
  example, the Quasar~3C175~(\copyright 1996 by NRAO). }
\label{fig:4}
\end{figure}

Jet engines in active galactic nuclei can be regarded as MHD generators filled
with a magnetised relativistic plasma [16].  In a rather simple though highly
instructive approach an extragalactic jet can be considered as giant
manifestation of an electric circuit~(Fig.~\ref{fig:4}).  The AGN-MHD generator plays
the role of the power/voltage supply.  The hot spot region represents the load.
The current must flow along the jet in not ideal conducting ``wires'', since
jets reveal themselves by the synchrotron emission from the radio to X-ray
regime.  The radiating electrons that carry the jet current suffer synchrotron
energy losses and thus, need to be re-accelerated all along the jet.  The
re-acceleration must be caused by electric fields directed along the current
flow.  To be more specific in terms of an electric circuit description one deals
with an electrical inductance, i.e.~one faces time varying currents caused by
the radiative dissipation.  In collisionless plasmas a time varying current
density is connected via Ohm's law to a collisionless electrical conductivity
due to particle inertia [17, 18].  It is this kind of conductivity $\Sigma$ that
relates the field-aligned electric field to the field-aligned electric current
density $j_\parallel$ in the case of extragalactic jets
\begin{equation}
E_\parallel = \frac{m_{\rm e}}{n e^2} \left| \Dif{\Vec{j}}{t}
  + \Nab \cdot ( \Vec{vj} + \Vec{jv}) \right|_\parallel =\Sigma j_\parallel
=\frac{\lambda_{\rm
  skin}^2}{L_{\rm shear}^2}\frac{v_{\rm shear}}{c} B_\perp
\label{equ:0}
\end{equation}
with $\lambda_{\rm skin}=c(m_{\rm e}/4\pi ne^2)^{1/2}$ ($m_e$ and $n$ are the
electron mass and particle density), $L_{\rm shear}$ and $B_\perp$ denoting the
electron skin length, the characteristic length scale of the the magnetic shear,
and the strength of the toroidal magnetic field component caused by the
generator shear flow $v_{\rm shear}$, respectively.  Recently, we have shown
that in this scenario the in situ acceleration of electrons up to \Unit{T\eV}
energies is possible provided that magnetic reconnection taps the jet magnetic
field as the relevant energy source [19, 20] (for a discussion of magnetic
reconnection in the context of jets see also [21, 22]).  Since particle inertia
operates on relatively small spatial scales the jet magnetic field has to be
organised in a highly filamentary structure.  Observations indicate [e.g.~5, 23]
that indeed the jets are characterised by highly filamentary current-carrying
helical magnetic field configurations.  A characteristic feature of current
filamentation is the local reversal of the average current direction.  Thus,
besides leptons, in magnetic field-aligned electric fields $E_\parallel$ caused
by particle inertia protons should be efficiently accelerated towards the hot
spots.

The main question that concerns us here is to what maximum energies part of the
jet protons can be accelerated.  The $E_\parallel$ can be estimated from
Eq.~(\ref{equ:1}) as a fraction $\alpha=\frac{\lambda_{\rm skin}^2}{L_{\rm
shear}^2}\frac{v_{\rm shear}}{c}$ of the toroidal magnetic field.  The maximum
$E_\parallel$ appears in thin current filaments of the order of the electron
skin length which implies $\alpha=1$.  The following calculations are performed
for $\alpha=1$ and can be re-scaled straightforward for weaker electric fields.
The Lorentz factor $\Gamma_{\rm p}$ cosmic ray protons can gain in a linear jet
accelerator is limited by
\begin{equation}
  \Gamma_{\rm p} = \frac{e E_\parallel L_{\rm acc}}{m_{\rm p} c^2} , \label{equ:1}
\end{equation}
where $m_{\rm p}$ and $c$ are the proton mass and the speed of light.  The above
estimation is highly idealised, since in real jets the twisted and tangled
magnetic fields give an equally distributed $E_\parallel$ and even local field
reversals.  The net energy gain is given by the average of a stochastic
acceleration within the jet.  However, the randomness should only influence the
flux of the accelerated hadrons and not their maximum energy achievable.

The acceleration length $L_{\rm acc}$ is given by $L_{\rm acc}= \min\{ L_{\rm
Jet}$, $L_{\rm loss}\}$ where $L_{\rm Jet}$ denotes the extension of the jet and
the loss length $L_{\rm loss}$ depends on the governing loss mechanism,
i.e.~proton synchrotron radiation, inverse Compton scattering (${\rm p} + \gamma
\rightarrow {\rm p} + \gamma'$), pair production (${\rm p} + \gamma \rightarrow
{\rm p} + {\rm e}^+ + {\rm e}^-$) or photo pion production (${\rm p} + \gamma
\rightarrow \pi^+ + {\rm n}$; ${\rm p} + \gamma \rightarrow \pi^0 + {\rm p}$),
respectively.  The loss lengths of the first two mentioned processes are
comparable, if $U_{\rm Rad} = B^2/8\pi$ ($U_{\rm Rad}$: energy density of
radiation), which just means equipartition between radiation and fields.  The
loss lengths then read [24] $L_{\rm loss}^{\rm syn} = L_{\rm loss}^{\rm ic} = 6
\pi m_{\rm p}^3 c^2 / \sigma_{\rm T} m_{\rm e}^2\Gamma_{\rm p} B^2$ where
$\sigma_{\rm T}$ is the Thomson cross section.  The loss lengths for pair and
photo pion production are given by $L_{\rm loss}^{\rm pair}=(\kappa^{\rm
pair}\sigma^{\rm pair} n_\gamma)^{-1}$ and $L_{\rm loss}^{\rm pion}=(\kappa^{\rm
pion}\sigma^{\rm pion} n_\gamma)^{-1}$ where $\sigma^{\rm pair}$, $\sigma^{\rm
pion}$, $\kappa^{\rm pair}$, $\kappa^{\rm pion}$, and $n_\gamma$ represent the
cross sections against pair and photo pion production the fractional proton
energy losses during one interaction process, and the densities of the involved
photons.  The cross section and fractional energy loss per collision are given
by [25]
\begin{eqnarray}
  \sigma^{\rm pair} &=& a r_0^2 \left[ \frac{28}{9} \ln \left(
  \frac{2\epsilon_\gamma^{\rm p.r.}}{m_{\rm e} c^2} \right) - \frac{218}{27}
  \right] \nonumber\\
  \kappa^{\rm pair} &=& \frac{2 m_{\rm e}}{m_{\rm p}} ,
\label{equ:2}
\end{eqnarray}
where $r_0$, $a$, and $\epsilon_\gamma^{\rm p.r.}$ are the classical electron
radius, the fine-structure constant, and the photon energies in the proton rest
frame.  The asymptotic value for the cross section for very high
$\epsilon_\gamma^{\rm p.r.}$ is given by $\sigma^{\rm pair} \sim 10^{-26}
\Unit{cm}^2$.  Depending on the magnetic field strength and the frequency of the
involved photons at very high proton energies the photo pion losses may become
important [26]
\begin{eqnarray}
  \sigma^{\rm pion} &=& 7 \cdot 10^{-36} \Unit{cm}^2 \Unit{\eV}^{-1}
  (\epsilon_\gamma^{\rm p.r.} - 160 \Unit{M\eV}) \nonumber\\
  \kappa^{\rm pion} &=& \frac{\epsilon_\gamma^{\rm p.r.}}{m_{\rm p} c^2}
  \frac{1+m_{\rm pion}^2 c^2 / 2 \epsilon_\gamma^{\rm p.r.} m_{\rm p}}{1 + 2
  \epsilon_\gamma^{\rm p.r.}/ m_{\rm p} c^2}
\label{equ:3}
\end{eqnarray}
and the asymptotic value for the cross section $\sigma^{\rm pion} \sim 10^{-28}
\Unit{cm}^2$ for very high $\epsilon_\gamma^{\rm p.r.}$.

In absence of nearby powerful external photon sources, the loss processes have
to be examined for the microwave background photons and, more important, for the
observed radio, optical and X-ray synchrotron photons emitted by the electrons
re-accelerated in the jets.  The thermal relic photons do not hinder UHECR
acceleration in extragalactic jets.  For the relevant energy ranges $E_{\rm p}
\approx 10^{18}- 10^{21} \Unit{\eV}$ (at lower \Unit{\eV}'s the energy of the
relic photons $\epsilon_\gamma^{\rm p.r.}$ is not sufficient for pair
production) the proton mean free path is $L_{\rm loss} = \min\{ (\sigma^{\rm
pair}\kappa^{\rm pair} n_\gamma)^{-1}, (\sigma^{\rm pion}\kappa^{\rm pion}
n_\gamma)^{-1} \} \ge 15 \Unit{Mpc}$ given [26, 27] a relic photon density of
$n_\gamma\approx 400 \Unit{cm}^{-3}$.  Thus, the proton mean free paths against
pair and photo pion production via thermal background photons largely exceed
typical jet lengths (synchrotron as well as inverse Compton losses are
negligible in this case).  On the other hand, if protons can be accelerated in
extragalactic jets up to the highest observed cosmic ray energies, the jets
should not be further away than $\sim 15 \Unit{Mpc}$ in order to contribute
significantly to the UHECR-flux detected on earth.

Equipartition of the magnetic energy density and the photon energy density
photons is a consequence of the synchrotron emission of the jet leptons.  The
number density of the synchrotron photons that reveal the existence of the
relativistic extragalactic jets is given as a function of the Lorentz factor of
the radiating electrons and the magnetic field strength $ n_\gamma=B^2/4 h
\Gamma_{\rm e}^2\omega_{\rm c}$ where $h$ and $\omega_{\rm c} = e B / m_{\rm e}
c$ are the Planck constant and the electron gyro frequency.  Making use of the
relation between the frequency $\nu$ of the observed synchrotron emission and
the electron Lorentz factor $\nu = \Gamma_{\rm e}^2eB/2\pi m_{\rm e}c$ one can
determine $\Gamma_{\rm e}$ and therefore $n_\gamma = B^2/8\pi h\nu$.  The
observationally determined magnetic field strengths in jets are of the order [5,
23] of some $1-10 \Unit{\mu G}$.  The synchrotron radiation is caused by
electrons with Lorentz factors $\Gamma_{\rm e}\sim 10^3-10^8$.

If the loss lengths are shorter than the length of the jets, the achievable
Lorentz factors for the protons accelerated within extragalactic jets can be
calculated from Eq.~(\ref{equ:1}) and the expression for the photon density
$n_\gamma$.  When pair and photo pion production dominate synchrotron/inverse
Compton losses, we receive
\begin{equation}
  \Gamma_{\rm p} = \frac{8 \pi e h}{m_{\rm p} c^2 \kappa \sigma}
  \frac{\nu}{B}. \label{equ:4}
\end{equation}
Here $\kappa$ and $\sigma$ denote the transfered energy rates per collision and
cross sections for either the pair or the photo pion production process,
respectively.  It should be noted, that $\kappa$ and $\sigma$ might depend on
$B$ due to $\epsilon_\gamma^{\rm p.r.}$.  For the second case, when pair and
photo pion production losses are negligible against synchrotron/inverse Compton
losses, one finds
\begin{equation}
  \Gamma_{\rm p} = \frac{m_{\rm p}}{m_{\rm e}} \sqrt{\frac{6\pi e c}{\sigma_{\rm
  T}B}}. \label{equ:5}
\end{equation}

If we assume $B \approx 10 \Unit{\mu G}$, the high frequency X-ray photons
($\sim \Unit{k\eV}$) result in proton energy losses via pair production starting
from $\Gamma_{\rm p} \approx 10^3$.  Given a threshold energy for the $p +
\gamma \rightarrow \pi^+ + n$ process of $\epsilon_0=160 \Unit{M\eV}$ proton
energy losses by photo pion production start from $\Gamma_{\rm p} \approx
10^{5}$.  However, the number density of the non-thermal photons available for
the proton energy loss processes depend on $\Gamma_{\rm e}^{-2}$.  The density
of these high energy photons is too small to effectively slow down the protons
via photo pion production.  We can state that the X-ray photons do not limit the
proton acceleration.

\begin{figure}
  \mbox{\includegraphics[width=0.49\textwidth]{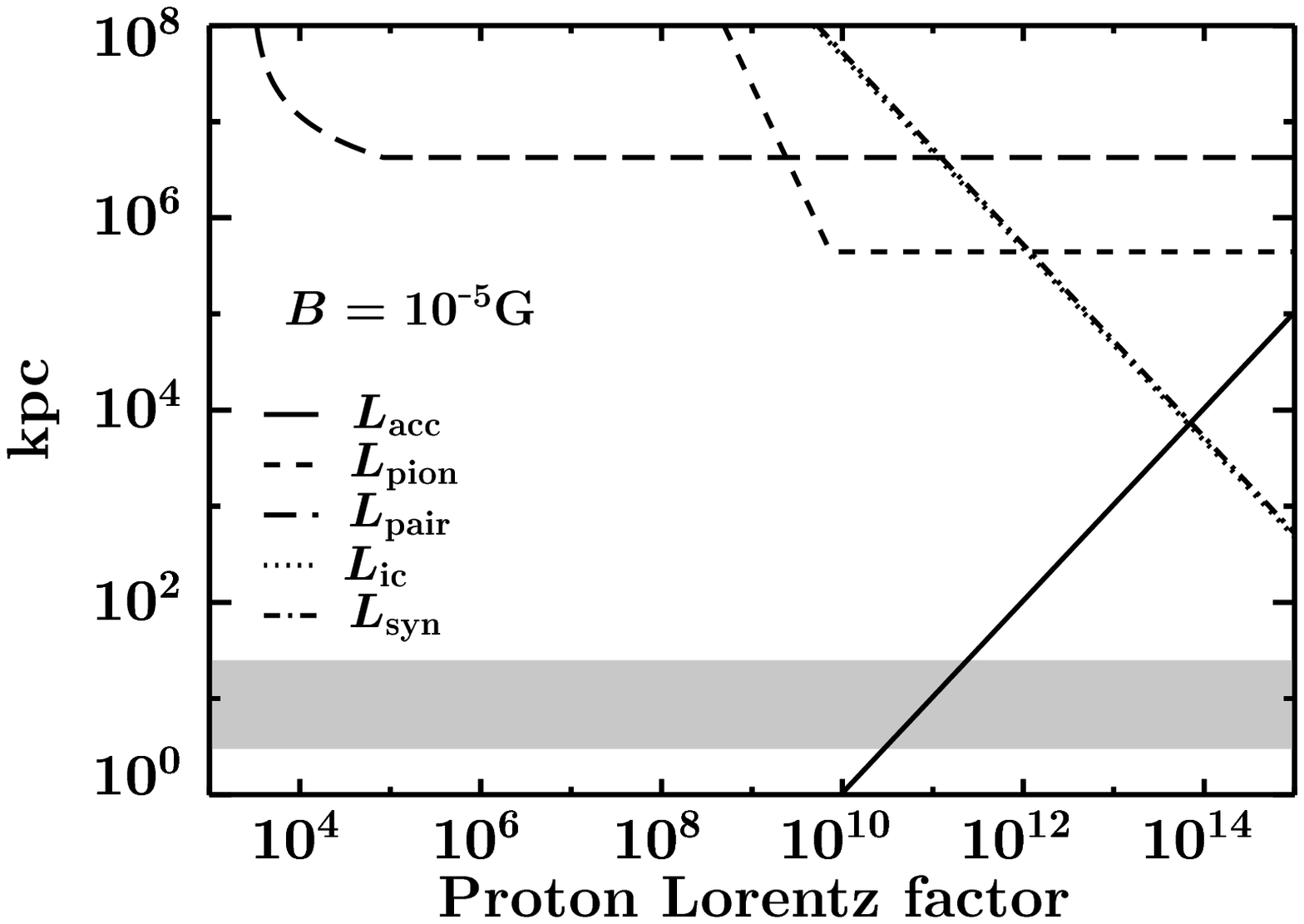}\hfill
  \includegraphics[width=0.49\textwidth]{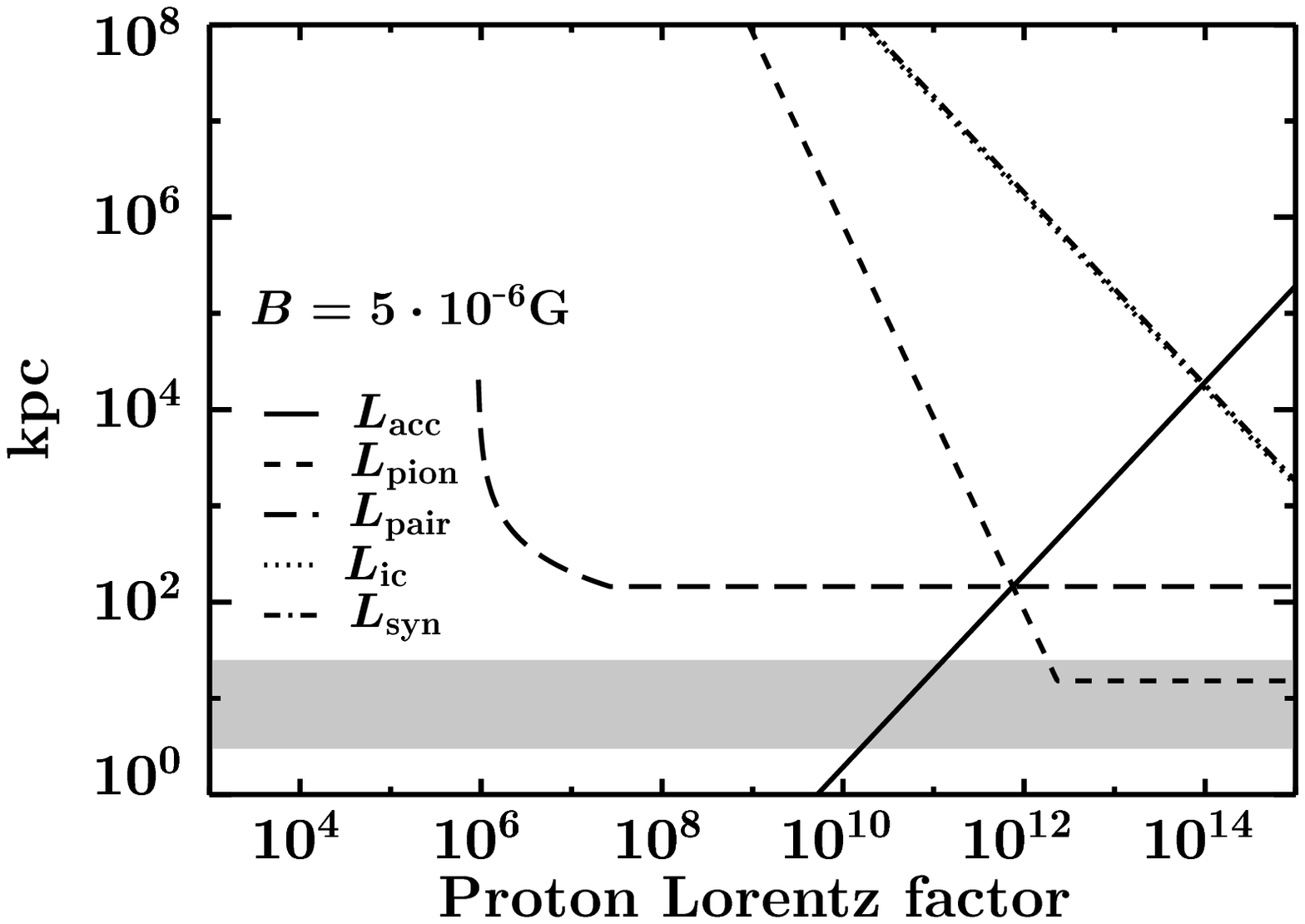}}
  \caption{The dependence of the loss lengths and the acceleration length on the
 	proton Lorentz factor is displayed.  The hatched area represents the range of
 	typical jet lengths.  In Fig.~\ref{fig:fig.1/2}a the influence of optical
 	synchrotron photons is shown, whereas Fig.~\ref{fig:fig.1/2}b represents the
 	effect of radio synchrotron emission.  The displayed lengths depend on the
 	magnetic field strength also.  Here, $B=10 \Unit{\mu G}$
 	(Fig.~\ref{fig:fig.1/2}a) and $B=5 \Unit{\mu G}$ (Fig.~\ref{fig:fig.1/2}b)
 	have been chosen, for example.}
  \label{fig:fig.1/2}
\end{figure}

In Fig.~\ref{fig:fig.1/2} the loss lengths and the acceleration lengths (see
Eq.~(\ref{equ:1})) are displayed in double-logarithmic representation as
functions of the proton Lorentz factors for optical (Fig.~\ref{fig:fig.1/2}a,
$\nu=10^{14} \Unit{Hz}$), and radio (Fig.~\ref{fig:fig.1/2}b, $\nu=10^9
\Unit{Hz}$) synchrotron photons.  The grey areas show typical jet lengths for
comparison.  In Fig.~\ref{fig:fig.1/2}a the magnetic field strength is chosen as
$B=10 \Unit{\mu G}$ and in Fig.~\ref{fig:fig.1/2}b as $B=5\Unit{\mu G}$, for
example.  As for the x-ray photons the optical synchrotron radiation does not
limit the proton acceleration, i.e.~the principal acceleration length $L_{\rm
acc}$ is limited, in this case by the synchrotron/inverse Compton loss length
$L_{\rm syn, ic}$, far beyond the extension lengths of extragalactic jets
$L_{\rm Jet}$.  In the case of the radio emission for a magnetic field strength
of $B=5 \Unit{\mu G}$ the loss lengths for pair and photo pion production and
the acceleration length meet at the same Lorentz factor of $\Gamma_{\rm
p}\approx 8\cdot 10^{11}$.  In fact, protons could be accelerated up to this
energy in unusually long jets (for the longest known extragalactic jet, Pictor A
at a distance of $140 \Unit{Mpc}$, one finds $L_{\rm Jet}\approx 300
\Unit{kpc}$).

\begin{figure}
 	\includegraphics[width=0.49\textwidth]{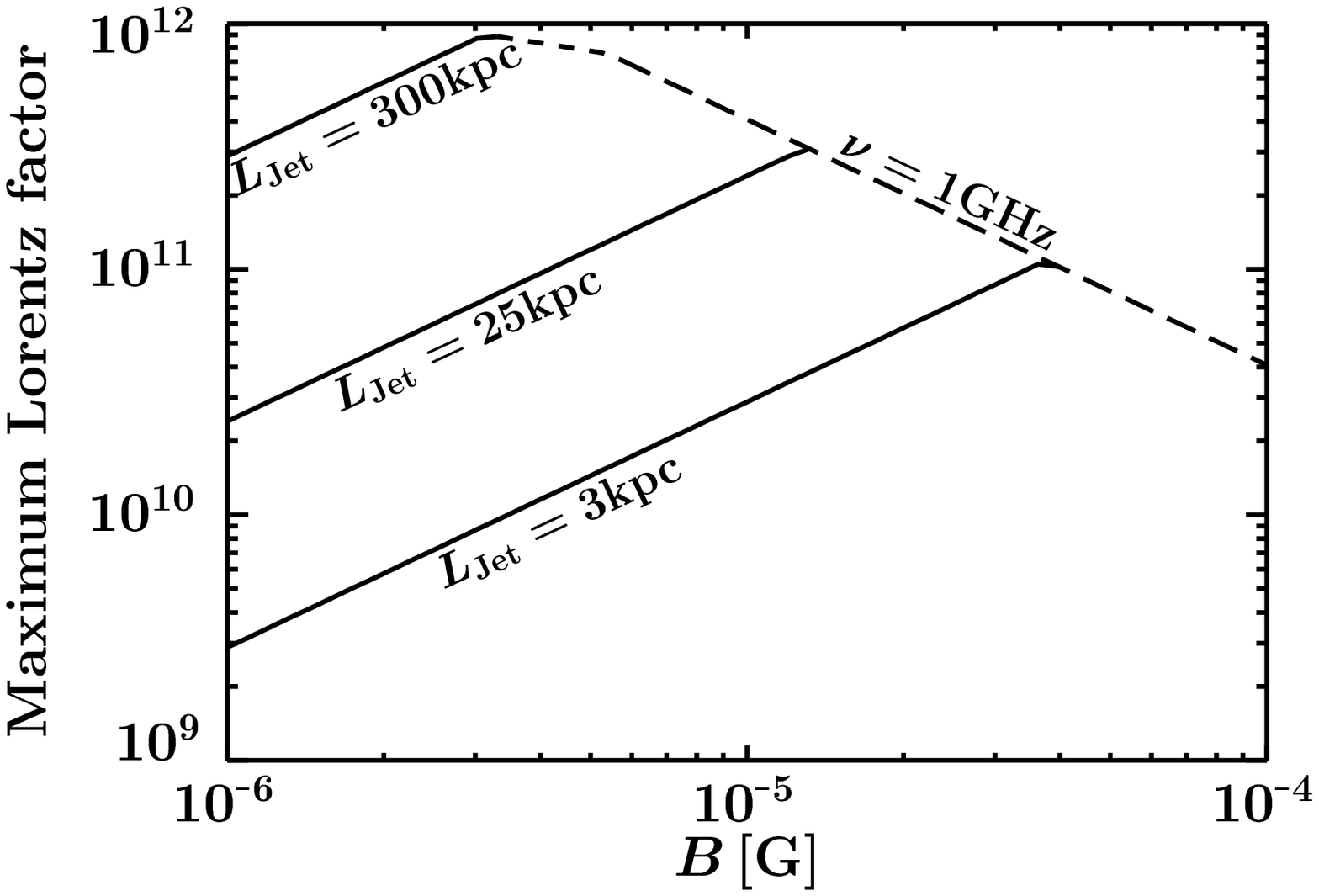}
  \caption{The maximum achievable Lorentz factor of UHECR protons as a function
  of the jet magnetic field strength ($B$) for different jet lengths ($L_{\rm
  Jet}$) and \Unit{GHz} radio synchrotron emission.}
  \label{fig:fig.3}
\end{figure}

The stronger the magnetic fields are the smaller are the loss lengths.  This is
due to the strong dependence $n_\gamma \sim B^2$.  On the other hand, stronger
fields imply higher field strengths available for particle acceleration, which
leads to a higher energy gain per unit length.  The maximum Lorentz factor,
protons can be accelerated to, depends on the magnetic field strength as shown
in Fig.~\ref{fig:fig.3} for L$_{\rm Jet}= 3 \Unit{kpc}$ to $L_{\rm Jet}= 300
\Unit{kpc}$ and $\nu=10^9 \Unit{Hz}$.  For relatively weak magnetic fields the
proton energy is limited by only the jet lengths (solid line).  Photo pion
production (short dashed line) gets important for very long jets (upper plot).
It is pair production that effectively limits the achievable Lorentz factor of
UHECR protons in typical extragalactic jets.  For example, for $L_{\rm Jet}=25
\Unit{kpc}$, which, in fact, is the extension of the Centaurs A jet and a field
strength of $B\approx 10 \Unit{\mu G}$ the pair production allows for
$\Gamma_{\rm p}\approx 3\cdot 10^{11}$.  It should be stressed that observations
indicate that, indeed, the Centaurus A jet has a magnetic field [4] of some $10
\Unit{\mu G}$.  Given an isotropic proton distribution injected in the jet only
a small fraction of protons with can gain the highest energies observed in the
UHECR population (a quantitative analysis of this point is possible in the
framework of collisionless magnetic reconnection [28]).
\section{Discussion}
The observed high energy synchrotron emission of extragalactic radio jets in the
optical and X-ray-regime clearly indicates the presence of magnetic
field-aligned electric fields along the whole jet lengths.  The leptons need
such electric field to be continuously re-accelerated, otherwise they would
loose their \Unit{T\eV} energies on length scales that are orders of magnitudes
smaller than the observed jet lengths.

The hadronic jet components should be accelerated in these magnetic
field-aligned electric fields caused by the temporal variations of the electric
current density which flows along the jet axis.  Such current driven jets are
produced by the rotating magnetospheres of supermassive accreting black holes in
the central regions of active galactic nuclei.  The differential rotation of the
accretion disk induces currents along the rotation axis and in the ideal
magnetohydrodynamical case, where the electrical conductivity is infinitely
high, an electric field which is perpendicular to the magnetic field is the
consequence.  However, such a fully ideal MHD-jet cannot explain any kind of
dissipation like the observed particle acceleration.  Some deviation from
idealness, i.e.~a localised finite electrical conductivity, is required on order
to explain the nonthermal jet radiation.  It is a well-known fact, that nonideal
processes in plasmas are always related to field-aligned electric fields which,
of course, are perfect particle accelerators.  In other words some part of the
perpendicular electric fields generated by the AGN-disk interaction can be
tapped for direct particle acceleration along the jet depending on the plasma
conductivity.  In the proposed scenario it is particle inertia that gives rise to
the violation of ideal Ohm's law.  Since the inertia effect only becomes
important on relatively small spatial scales, filamentary currents are
indispensable for the suggested model.  Also, in thin current filaments only
relatively strong field-aligned electric fields are reasonably to be expected.
In fact, one should bear in mind that for our calculations we used that
strongest $E_\parallel$ possible, namely $E_\parallel=\alpha B_\perp$ with
$\alpha=1$.  Our findings have to be re-scaled accordingly for $\alpha<1$.  The
Lorentz factors the protons can gain in these electric fields are mainly
determined in a quite complex way by the strength of the jet magnetic field and
the synchrotron photon background.  For typical jet parameters protons, in
principle, can be energised up to $\approx 10^{20} \Unit{\eV}$.  Our model
considers all relevant loss processes and therefore clearly shows its own
limitations.  The balance of the suggested acceleration mechanism and the
respective dominant loss process that gives an upper limit for the proton energy
may help to explain the observed energy range of the measured highest energy
UHECR particles in a promising manner.  This means it seems understandable not
only how, why and under what circumstances protons can gain the measured
energies, but also why we do not see particles with higher energies.

The magnetic field strength is the strongest constraint for possible UHECR
source candidates.  Magnetic fields significantly lower than $\approx 10
\Unit{\mu G}$ are not sufficient to accelerate hadrons to the highest observed
energies.  On the other hand, if magnetic fields are significantly stronger than
some $10\mu$G the synchrotron photon densities and thus the losses are too high
to allow for extremely high Lorentz factors.  We find, that efficient
acceleration is limited to magnetic fields very close to $\approx 10 \Unit{\mu
G}$, for the typical jet lengths of $3$--$25$~\Unit{kpc} found within a distance
of $15 \Unit{Mpc}$.  Consequently, we feel that the Centaurus A jet, located
$3.4 \Unit{Mpc}$ away, should significantly contribute to the observed UHECR
population at the highest energy levels.  According to our calculations
Centaurus A may even be able to produce cosmic rays with energies up to the
presently still unique Fly's Eye event [15] of $3.2 \cdot 10^{20} \Unit{\eV}$.
\section{Acknowledgements}
This work has kindly been supported by the Deutsche Forschungsgemeinschaft
through the grant {LE 1039/4-1}.
\clearpage

\clearpage
\begin{thebibliography}{}
\bibitem{tur97} T.J.~Turner, I.M.~George, R.F.~Mushotzky, K.~Nandra, {\em Astrophys.~J.\/} {\bf 475} (1997) 118.
\bibitem{isr98} F.P.~Israel, {\em Astron.~Astrophys.~Rev.\/} {\bf 8} (1998) 237.
\bibitem{roe00} H.-J.~R{\"o}ser, K.~Meisenheimer, M.~Neumann, R.G.~Conway, R.A.  Perley {\em Astron.~Astrophys.}, {\bf 360} (2000) 99.
\bibitem{bur83} J.O.~Burns,  E.D.~Feigelson, E.J.~Schreier, {\em Astrophys.~J.\/} {\bf 273} (1983) 128.
\bibitem{mei96} K.~Meisenheimer, H.-J.~R{\"o}ser, M.~Schl{\"o}telburg, {\em Astron.~Astrophys.\/} {\bf 307} (1996) 61.
\bibitem{cel93} A.~Celotti, A.C.~Fabian, {\em Mon.~Not.~Roy.~Astr.~Soc.\/} {\bf 624} (1993) 228.
\bibitem{war98} J.F.~Wardle, D.C.~Homan, R.~Ojha, D.H.~Roberts, {\em Nature\/} {\bf 395} (1998) 457.
\bibitem{hom99} D.C.~Homan, J.F.~Wardle, {\em Astron.~Journal\/} {\bf 118} (1999) 1942.
\bibitem{sik00} M.~Sikora, G.~Madejski, {\em Astrophys.~J.\/} {\bf 534} (2000) 109.
\bibitem{beg84}  M.C.~Begelman, R.D.~Blandford  M.J.~Rees, {\em Rev.~Mod.~Phys.} {\bf 56} (1984) 255.
\bibitem{ver96} R.C.~Vermeulen, {\em Lect.~Notes Phys.} {\bf 471} (1996) 245.
\bibitem{ben78} G.~Benford, {\em Mon.~Not.~Roy.~Astr.~Soc.} {\bf 183} (1978) 29.
\bibitem{vil94} M.~Villata, A.~Ferrari, {\em Astron.~Astrophys.\/} {\bf 284} (1994) 663.
\bibitem{lesch89} H.~Lesch, S.~Appl, M.~Camenzind, {\em Astron.~Astrophys.\/} {\bf 225} (1989) 341.
\bibitem{bir93} D.J.~Bird, et al., {\em Phys.~Rev.~Lett.\/} {\bf 71} (1993) 3401.
\bibitem{blandf90}  R.D.~Blandford, in Active Galactic Nuclei, eds.~T.J.L.~Courvosier and M.~Mayor, Saas-Fee Advanced Course 20, Springer, Berlin (1990) 161.
\bibitem{spei70} T.W.~Speiser, {\em Planet.~Space Sci.\/} {\bf 18} (1970) 613.
\bibitem{lyo85} L.R.~Lyons, T.W.~Speiser, {\em J.~Geophys.~Res.\/} {\bf 90} (1985) 8543.
\bibitem{les98} H.~Lesch, G.T.~Birk,  {\em Astrophys.~J.\/} {\bf 499} (1998) 167.
\bibitem{bir00}  G.T.~Birk, H.~Lesch, H., {\em Astrophys.~J.\/} {\bf 530} (2000) L77.
\bibitem{bla96}  E.G.~Blackman, {\em Astrophys.~J.\/} {\bf 456} (1996)  L87.
\bibitem{lit99} Y.E.~Litvinenko, {\em Astron.~Astrophys.\/} {\bf 349} (1999) 685.
\bibitem{bir96} J.A.~Biretta, in Solar and Astrophysical Magnetohydrodynamic Flows, ed.~by K.C.~Tsinganos, Kluwer, Dordrecht (1996) 357.
\bibitem{ryb79} G.B.~Rybicki, A.P.  Lightman, Radiative Processes in Astrophysics, Wiley, New York (1979)
\bibitem{lon81} M.S.~Longair, High energy astrophysics, Cambridge University Press (1981)
\bibitem{ber88} V.S.~Berezinsky, S.I.~Grigor'eva, {\em Astron.~Astrophys.\/} {\bf 199} (1988) 1.
\bibitem{blu70} G.~Blumenthal, {\em Phys.~Rev.~D\/} {\bf 1} (1970) 1596.
\bibitem{vek97} G.E Vekstein, P.K.~Browning, {\em Phys.~Plasmas} {\bf 4} (1997) 2261.

\end{thebibliography}
\end{document}